\documentclass[11pt]{article}

\usepackage{amssymb}
\usepackage{amsmath}
\usepackage{hyperref}
\numberwithin{equation}{section}

\begin{document}

\title{The Clifford Algebra approach to Quantum Mechanics A: The Schr\"{o}dinger and Pauli Particles.}
\author{B. J. Hiley\footnote{E-mail address b.hiley@bbk.ac.uk.}\hspace{0.1cm}  and R. E. Callaghan.}
\date{TPRU, Birkbeck, University of London, Malet Street, \\ London WC1E 7HX. }
\maketitle

\begin{abstract}
In this paper  we show how all the quantum properties of Schr\"{o}dinger and Pauli particles can be described entirely from within a Clifford algebra taken over the reals.  There is no need to appeal to any `wave function'.  To describe a quantum system, we define the Clifford density element [CDE],  $\rho_{c}=\Phi_{L}\widetilde\Phi_{L}$, as a product of an element of a minimal left ideal, $\Phi_{L}$, and its Clifford conjugate.   The properties of the system are then completely specified in terms of bilinear invariants of the first and second kind calculated using the CDE. Thus the quantum properties of a system can be completely described from within the algebra without the need to appeal to any Hilbert space representation.

Furthermore we show that the essential bilinear invariants of the second kind are simply the Bohm energy and the Bohm momentum, entities that make their appearance in the Bohm interpretation.  We also show how these parameters emerge from standard quantum field theory in the low energy, single particle approximation. There is no need to appeal to classical mechanics at any stage. This clearly shows that the Bohm approach is entirely within the standard quantum formalism.  The method has enabled us to lay the foundations of an approach that can be extended to provide a complete relativistic version of Bohm model.  In this paper we confine our attention to the details of the non-relativistic case and will present its relativistic extension in a subsequent paper.
\end{abstract}

\section{Introduction}


\subsection{The General Principles}


The methods that we use in this paper developed out of a more ambitious enquiry into the nature of quantum space-times.  Although the details of this exploration are not necessary for what follows, we feel that it would be helpful to explain the context in which they arose.

Since the advent of general relativity in which matter and geometry codetermine each other, there is a growing realisation that starting from an {\em a priori} given space-time manifold in which we allow material processes to unfold is, at best, limited. Can we start from something more primitive from which both geometry and material process unfold together?  The challenge is to find a formalism that would allow this to happen. 

 In the early sixties David Bohm  introduced the notion of a discrete structural process, \cite{db65}, \cite{db71}, in which he takes as basic, not matter or fields in interaction {\em in} space-time, but a notion of `structure process'  from which the geometry of space-time and its relationship to matter emerge together providing a way that could underpin general relativity and quantum theory. Bohm provides a detailed discussion of the general notions implicit in this approach, but the problem of how  to develop these ideas into a well defined  mathematical structure remained unanswered.  
 
  Since these attempts were first made, there has been a considerable amount of mathematical work exploring a possible deeper structure to space-time of which Bohm was unaware.  These attempts have now become better known to the physics community under the term `non-commutative geometry' and there exists an excellent review of the  evolution of these ideas by Pierre Cartier \cite{pc01}.  Our own thinking has been very influenced by this pioneering work together with the work of  Gel'fand  \cite{dhl97}, \cite{jv06} and Allan Connes \cite{ac90}.

 Our specific interest is to explore some of the possible implications of this work for quantum physics in the light of Bohm's early proposals.   In a recent paper, one of us [BJH]  \cite{bh08} has shown that by starting from a general notion of structure process, and without assuming a space-time background, we are led to a Clifford groupoid as a way of ordering the rotational symmetries of  quantum processes. This gives us a simple example of how we can describe the order of a discrete structure process. By adding more structure, we are able to arrive at the full orthogonal Clifford algebra.  We have also proposed a symplectic Clifford structure \cite{ac} which is essential to complete our project, but here we will confine our attention to the orthogonal Clifford.  
 
This result is not unexpected because Clifford's original motivation was to develop a process based approach to {\em classical non-relativistic geometry}, a point that has been emphasised by Hestenes \cite{dh03} and by Doran and Lasenby \cite{dl}. But we know that the Clifford algebra can be extended to include relativity and plays an essential role in the Dirac theory of the relativistic electron.  There the elements of the algebra are treated as operators which then act on an external Hilbert space.  In this paper we will show that this external Hilbert space is not necessary, as everything can be done within the algebra itself.  The relevance of this comment is to support out claim that everything can be abstracted from a structure process described by an algebra.

Most of the work in \cite{bh08} was to establish how it is possible to produce  an algebraic description of this structure process.  Having demonstrated how this was possible, we went on to show that this algebra had enfolded in it a series of what we called  `shadow manifolds'. We deliberately choose the plural `manifolds' because we have a non-commutative algebra and as shown by Gel'fand's approach, we can only abstract a unique manifold if our algebra is commutative.

 Notice that our approach stands the conventional approach on its head, as it were, because we start with the algebra and then abstract the geometry. We do not start with a {\em a priori} given manifold and then build an algebra on that.  The philosophical motivation for starting in this way comes from Bohm's notion of the implicate and explicate order \cite{db80}.  The process described by the non-commutative algebra is the implicate order, while the shadow manifolds are the explicate orders.   Rather than having a single Minkowski space-time manifold,  we now have many shadow space-times, each corresponding to what is usually called a `Lorentz frame'.  Thus each shadow manifold corresponds to an equivalence class of Lorentz observers.

However this is all very classical and the discussion so far has little to do with anything quantum.  We are still in the classical domain even though our algebras contain spinors, these spinors actually describe light rays and these are ordered to give a classical light ray geometry as has been extensively discussed by Penrose and Rindler \cite{rpwr}.  No notion of quantum theory has yet been introduced so how then do the quantum aspects of these algebras appear?

\subsection{Quantum Mechanics in terms of Clifford Algebras.}


In the final section of Hiley \cite{bh08}, we outlined very briefly a general method to describe all the usual quantum properties of the Schr\"{o}dinger, Pauli and Dirac particles within the Clifford algebra structure itself without the need to make an appeal to Hilbert space.  The purpose of this paper is to show  in detail how this works for the Schr\"{o}dinger and  Pauli theories.  We will leave the Dirac theory to a separate paper since it introduces several additional features that need careful discussion \cite{bhbc081}. 

In \cite{bh08} we applied the Clifford algebra ${\cal{C}}_{(0,1)}$ to the Schr\"{o}dinger particle.  We started by constructing an element of the minimal left ideal and showed how it encoded all the information normally contained in the conventional wave function (for more details see the earlier work of Frescura and Hiley \cite{ffbh81}).  Thus there is no need to use the concept of a wave function.  This means we do not introduce `quantum  operators' representing `observables' acting on wave functions.  Instead the properties of the quantum process are obtained in terms of specific elements of the algebra itself, a point already made by Haag \cite{haa92}. It is the purpose of this paper to present the details of this method.

An immediate advantage of the algebraic approach is that it provides a natural mathematical  hierarchy in which to embed the physical hierarchy:-  a non-relativist particle without spin (Schr\"{o}dinger), a non-relativistic particle with spin (Pauli) and a relativistic particle with spin (Dirac).  These features fit naturally into the nested hierarchy,  ${\cal C}_{0,1}$ (Schr\"{o}dinger), ${\cal C}_{3,0}$ (Pauli), ${\cal C}_{1,3}$ (Dirac)\footnote{Here the suffixes correspond to the signature of the metric tensor of the shadow manifolds.}.  This advantage has already been noted by Hestenes and Gurtler \cite{dh71}.

 Since we confine our attention in this paper to the Schr\"{o}dinger and Pauli theories, we will consider only the Clifford algebras ${\cal C}_{0,1}$ and ${\cal C}_{3,0}$.   In section 2.1, we show how the elements of the left and right ideals are constructed in each case.  This enables us to construct bilinear invariants of the first kind in sections 2.2 and 2.3.  These are not sufficient to completely specify the state of the process so, in section 2.4, we construct bilinear invariants of the second kind \cite{TT57}.  It is these invariants that lead directly to the Bohm energy and momentum.  We then go on to construct  a Clifford bundle over a chosen shadow manifold\footnote{We hope this is a temporary expedient until we have fully understood the role of the symplectic Clifford algebra.}.  Then, in section 2.6, we replace the directional derivative on the manifold by {\em two} generalized Dirac derivatives on the bundle, $\overrightarrow D$ and $\overleftarrow D$, one acting on the left and the other acts on the right. These are then used to construct bilinear invariants which in turn are used to describe the time evolution of the quantum process.  
 
 In section 3 we discuss the Schr\"{o}dinger theory, leaving section 4 for a full discussion of the Pauli theory. Finally we compare and contrast our approach to the Pauli theory with that of Bohm, Schiller and Tiomno \cite{BST}.   
\section{The Basis of the Clifford Algebraic Approach}
 
 
\subsection {Elements of Minimal Left Ideal.}


The starting point of our approach is to introduce an element of a minimal left ideal, $\Phi_{L}({\boldsymbol r},t)$.  We do this by choosing a specific primitive idempotent, $\epsilon$, so that we can write $\Phi_{L}({\boldsymbol r},t)=\phi_{L}({\boldsymbol r},t)\epsilon$.  Here $\phi_{L}({\bf r}, t)$ is a linear combination of some of the elements of the algebra.  The choice of the particular form of $\epsilon$ is determined by the physical situation we are considering. The specific choice will be come clear as we consider particular physical examples.  

As explained in Hiley \cite{bh08}, the element $\phi_{L}({\boldsymbol r},t)$ contains all the information contained in the standard wave function.  However it must be stressed that although we use the symbol $\Phi_{L}$ to denote the {\em algebraic element}, it should not be confused with the symbol $\psi$ which is an element of an external {\em Hilbert space}.  When we come to find particular matrix representations of $\Phi_{L}$, we will find it is a square $n \times n$ matrix, whereas the vector in Hilbert space, $\psi$, is represented by an $n\times 1$ column matrix.  Under a rotation, both transform via the covering spin group, the element of the left ideal undergoing the two-sided transformation $g\Phi_{L}g^{-1}$, whereas the element of the Hilbert space, usually called a spinor\footnote{The use of the term `spinor' has caused and still causes confusion.  When we use the term, we apply it to the mathematical object that undergoes a transformation of the spin group. In this case both $\Phi_{L}$ and $\psi$ are spinors, albeit different representations of the same object.  Furthermore since we include a discussion of  the Schr\"{o}dinger case in a Clifford algebra, the ordinary Schr\"{o}dinger wave function is a spinor.  This is why we have remarked earlier that quantum processes ``live in" the covering space.}, undergoes the one-sided transformation $g^{\prime}\psi$.

We also need an element of a right ideal, $\Phi_{R}({\boldsymbol r},t) $, which is, in some sense, dual to $\Phi_{L}({\boldsymbol r},t) $. For the dual we need to choose one of the several conjugates that arise naturally in any Clifford algebra.  These are either involutions or anti-involutions \cite{ip}.  We choose the particular anti-involution known as the {\em Clifford conjugation}\footnote{ Formal definitions of these automorphisms will be found in Porteous \cite{ip}.}. To explain how this works, we first notice that any element of the Pauli Clifford algebra, for example, can be written in the form $A=S +V+B+P$ where $S$ is the scalar part, $V$ is the vector part, $B$ is the bivector, and $P$ is the pseudoscalar part.  The Clifford conjugation of $A$ is then defined by 
\begin{eqnarray}
\widetilde {A}= S-V-B+P						
\end{eqnarray}
In these terms we choose our element of the right ideal to be $\Phi_{R}({\boldsymbol r},t)=\widetilde\Phi_{L}({\boldsymbol r},t)$.

Actually we will find that we do not need to use elements of the left and right ideas separately.  Rather to specify the state of a quantum system we need the quantity defined by
\begin{eqnarray}
\rho_{c}({\boldsymbol r},t) =\Phi_{L}({\boldsymbol r},t) {\widetilde \Phi_{L}}({\boldsymbol r},t) 			
\end{eqnarray}
We will call $\rho_{c}$ the {\em Clifford density element} [CDE].  It plays a central role in our approach since it contains all the information necessary for us to describe the state of a system completely.   It plays a similar role to that of the usual density operator but, as we explained above, CDE is not an {\em operator} in the quantum sense.  It is simply a specific element in the algebra.
To put it another way, it `lives', as it were, in the Clifford algebra, and is not an operator on an external Hilbert space.  Nevertheless as we will show later, if we construct a matrix representation of this quantity, we find that it is isomorphic to the standard density matrix.  However it should be noted that the CDE, $\rho_{c}$,  is independent of any specific representation, as Hestenes \cite{dh03} has already emphasised.  This makes the overall mathematical structure of the theory much more transparent.

For simplicity, in this paper we will choose these CDEs to be idempotent which means that they describe what, in the usual approach, would be called `pure states'.  The CDEs can be extended to describe mixed states but we will leave a discussion of this extension to another paper where we will link these ideas to a general Moyal algebra \cite{bjh06}.

\subsection{Bilinear Invariants.}


As we have seen the coefficients $\phi_{L}$ are functions of $ \boldsymbol r$ and $t$, which are the co-ordinates on a specific shadow manifold determined by the projection $\eta$ as described in Hiley \cite{bh08} (see also below).  This means we have a CDE at each point of this shadow manifold.  Thus we have a Clifford algebra at each point of our shadow space-time manifold giving us a Clifford bundle. 

What we need to do now is to use the CDE to find the bilinear invariants that will characterise the physical properties of our quantum process.
To show how this works, let us recall that in standard quantum mechanics, the expectation value of operator $\hat{B}$ is found via:
\begin{eqnarray}
\langle \hat{B} \rangle =\langle \psi |\hat{B}| \psi \rangle =tr(\hat{B}\hat\rho) \nonumber
\end{eqnarray}
Here $\hat\rho$ is the standard density matrix.

To make contact with our algebraic approach we must replace the density matrix $\hat\rho$ by the CDE, $\rho_{c}$.  Then we use the following correspondences:-
\begin{eqnarray}
|\psi\rangle \rightarrow\Phi_{L}=\phi_{L}\epsilon\hspace{1cm}\langle\psi|\rightarrow \Phi_{R}={\widetilde\Phi_{L}}=\epsilon\phi_{R}=\epsilon{\widetilde\phi_{L}} \nonumber
\end{eqnarray}
and choose the algebraic equivalent to the operator $\hat{B}$.  Let us call this element $B$, then our expectation value becomes
\begin{eqnarray}
\langle B\rangle = tr(\epsilon{\widetilde\phi_{L}} B \phi_{L}\epsilon) \nonumber
\end{eqnarray}
Since the trace is invariant under cyclic permutations, we can arrive at the form
\begin{eqnarray}
\langle B\rangle = tr({\widetilde\phi_{L}} B \phi_{L}\epsilon)= tr(B\phi_{L}\epsilon{\widetilde\phi_{L}}) \label{eq:expt}
\end{eqnarray}							
But $ \phi_{L}\epsilon{\widetilde\phi_{L}}=\rho_{c}$, so we have
\begin{eqnarray}
\langle B\rangle = tr(B\rho_{c}) \label{eq:expt}
\end{eqnarray}							
However because we can write
\begin{eqnarray*}
B = b^{s} + \sum b^{i}e_{i} + \sum b^{ij}e_{ij} + \sum b^{ijk}e_{ijk}+ \dots
\end{eqnarray*}
 the mean value of any dynamical element, $B$, in the Clifford algebra becomes
\begin{eqnarray}
tr(B\rho_{c}) = tr(b^{s}\rho_{c} + \sum b^{i}e_{i}\rho_{c} +\sum b^{ij}e_{ij}\rho_{c} +\dots)\hspace{1.3cm}  \nonumber\\
= b^{s}tr(\rho_{c}) + \sum b^{i}tr(e_{i}\rho_{c}) + \sum b^{ij}tr(e_{ij}\rho_{c})/2 \dots\nonumber
\end{eqnarray}
This shows that the state of our system is specified by a set of bilinear invariants
\begin{eqnarray}
tr(1\rho_{c})\rightarrow\mbox{scalar}\hspace{0.7cm}tr(e_{j}\rho_{c})\rightarrow\mbox{vector}\hspace{0.7cm}tr(e_{ij}\rho_{c})\rightarrow\mbox{bivector}\hspace{0.7cm}tr(\dots)\rightarrow\dots\hspace{0.7cm} \nonumber 
\end{eqnarray}
These bilinear invariants then characterise the physical properties of the quantum process in the algebraic approach.

\subsection{Bilinear Invariants of the First Kind.}

The simplest bilinear invariants that can be formed take $B$ to be an element of the appropriate Clifford algebra.  We will follow Takabayashi \cite{TT57} and call $\langle B \rangle$, a bilinear invariant of the first kind. Thus in the particular case of the Dirac Clifford algebra, we can form 16  bilinear invariants. 
Unfortunately not all of these invariants are linearly independent. Takabayashi \cite{TT57} has shown that these invariants give only 7 independent real quantities\footnote{If normalisation is taken into account, then there are only 6 independent real quantities.} as opposed to the 8 real quantities we need to specify the state.  (Recall we have four complex parameters  in the standard Dirac spinor).  This reduction from 16 to 7 arises because there 9 independent kinematical identities among the invariants.  

In a similar way, it is not difficult to show that for the Pauli algebra there are only 3 independent real quantities, as opposed to the 4 required by the standard Pauli spinor.  In the case of the Schr\"{o}dinger algebra, there is only one real quantity whereas we clearly need 2 real quantities in the wave function.  Indeed it is quite easy to see what information is missing in each of these  cases.  The bilinear invariants of the first kind give no information about the phase.

\subsection{Bilinear Invariants of the Second Kind.}

In quantum theory the phase itself carries information about the momentum and the energy. To obtain expressions for the energy-momentum, we need to take derivatives like $\partial/\partial t$ and $\partial/\partial x$.  This suggests that we construct invariants involving such derivatives.  For the convenience of notational simplicity, we will denote these derivatives generically by $\partial$. Thus we can form invariants like
\begin{eqnarray}
(\partial\Phi_{L}){\widetilde\Phi_{L}} \pm \Phi_{L}(\partial{\widetilde\Phi_{L}}) 	\label{eq:meanpi}
\end{eqnarray}
If we choose the plus sign, we will simply have a derivation of a bilinear invariant of the first kind and that will clearly not give us any information about the phase.  Thus we must choose the negative sign, which we will write in a simplified notation as
\begin{eqnarray}
\Phi_{L}\overleftrightarrow\partial\widetilde\Phi_{L}=(\partial\Phi_{L}){\widetilde\Phi_{L}} - \Phi_{L}(\partial{\widetilde\Phi_{L}}) 	\label{eq:meanpi2}
\end{eqnarray}
Takabayashi \cite{TT57} call these bilinear invariants of the second kind.
Those  familiar with quantum field theory will immediately recognise this as the type of notation used for the standard energy-momentum tensor of quantum field theory.  

To bring this out clearly, let us examine the energy-momentum tensor for the Schr\"{o}dinger field.  Recall that in the standard approach, the Schr\"{o}dinger equation may be deduced from the Lagrangian
\begin{eqnarray*}
{\cal{L}}= - \frac{1}{2m}\nabla \psi^{*}\cdot\nabla \psi+\frac{i}{2}[(\partial_{t}\psi)\psi^{*}-(\partial_{t} \psi^{*})\psi]-V\psi^{*}\psi.
\end{eqnarray*}
Here $\psi$ without the suffix $L$ is the usual wave function.
The standard energy-momentum tensor is defined by
\begin{eqnarray*}
T^{\mu\nu}=- \left\{\frac{\partial {\cal{L}}}{\partial(\partial^{\mu}\psi)}\partial^{\nu}\psi+\frac{\partial {\cal{L}}}{\partial(\partial^{\mu}\psi^{*})}\partial^{\nu}\psi^{*}\right\}+\delta_{\mu\nu}{\cal{L}}
\end{eqnarray*}
so that the momentum density can be written as
\begin{eqnarray*}
T^{0j}=-\left\{\frac{\partial {\cal{L}}}{\partial(\partial^{0}\psi)}\partial^{j}\psi+\frac{\partial {\cal{L}}}{\partial(\partial^{0}\psi^{*})}\partial^{j}\psi^{*}\right\}.
\end{eqnarray*}
Since
\begin{eqnarray*}
\frac{\partial {\cal{L}}}{\partial(\partial^{0}\psi)}=\frac{1}{2i}\psi^{*}
\hspace{0.5cm}\mbox{and}\hspace{0.5cm}
\frac{\partial {\cal{L}}}{\partial(\partial^{0}\psi^{*})}=-\frac{1}{2i}\psi,
\end{eqnarray*}
we find
\begin{eqnarray*}
T^{0j}=\frac{i}{2}\left[\psi^{*}\partial^{j}\psi - \psi\partial^{j}\psi^{*}\right]
\end{eqnarray*}
with $\partial^{j}=-\nabla$.
We immediately recognise that $T^{0j}$ is the current, which is also the Bohm momentum, ${\bf p}_{B} ={\boldsymbol\nabla} S$, as can be easily seen to be the case, provided we write $\psi=R\exp[iS]$.  A similar argument goes through for the energy using
\begin{eqnarray*}
T^{00}=\frac{i}{2}[\psi^{*}\partial^{0}\psi-\psi\partial^{0}\psi^{*}].
\end{eqnarray*}
This gives the Bohm energy, $E_{B}=-\partial_{t}S$.

Thus the Bohm energy-momentum is nothing but the energy-momentum density derived from the standard expression for the energy-momentum tensor. We will show later in this paper that similar results follow for the Pauli theory.  This idea has already been exploited by Horton, Dewdney and Nesteruk \cite {ghcd00} for the Klein-Gordon equation. We find the same applies to the Dirac theory as we will show in a later paper \cite{bhbc081}.

This discussion shows that the Bohm energy-momentum density is not something {\em ad hoc} as claimed by Heisenberg \cite{wh} but contains vital information about the phase factor that is essential for a complete definition of the quantum process. Thus the Bohm energy-momentum density is not something that is imported from classical mechanics, it is already part of standard quantum field theory. This means that there is no need to make any appeal to classical mechanics whatsoever.    This avoids the questionable step in Bohm's \cite{db52} original argument of
 replacing the classical action by the phase of the wave function.  It also shows there is no need to derive the relation  ${\bf p}_{B} ={\boldsymbol\nabla} S$ from other assumptions as is proposed by D\"{u}rr, Goldstein and Zanghi \cite{dgz92}.

\subsection{General Algebraic Expressions for the Bohm Energy-Momentum.}
 
 The discussion of the previous subsection was in terms of the standard quantum formalism using wave functions. We must now re-express these formulae in terms of elements of the Clifford algebra. Therefore let us  define a momentum by
\begin{eqnarray}
\rho P^{j}(t)=-i\alpha\Phi_{L}\overleftrightarrow\partial^{j}\widetilde\Phi_{L}=-i\alpha\left [(\partial^{j}\Phi_{L}){\widetilde\Phi_{L}}-\Phi_{L}(\partial^{j}{\widetilde\Phi_{L}})\right]								
			\label{eq:P}					
\end{eqnarray}
and an energy by
\begin{eqnarray}
\rho E(t)=i\alpha\Phi_{L}\overleftrightarrow\partial^{0}\widetilde\Phi_{L}=i\alpha[(\partial^{0}\Phi_{L}){\widetilde\Phi_{L}}-\Phi_{L}(\partial^{0}{\widetilde\Phi_{L}})]
			\label{eq:E}					
\end{eqnarray}
Here $\alpha =1/2$ for the Schr\"{o}dinger case, while $\alpha = 1$ for the Pauli and the Dirac cases\footnote{ The reason for this difference is because the primitive idempotent for the Schr\"{o}dinger case is 1, whereas the Pauli and Dirac primitive idempotents are of the form $(1+\gamma)/2$ where $\gamma$ is an element of their respective algebras.  We discuss how these primitive idempotents are chosen later in this paper.}.  Notice that we do not yet attach a suffix $B$ to this momentum and energy.  This is because, although in the Schr\"{o}dinger case these expressions reduce to the Bohm momentum and energy, in the Pauli and Dirac cases they contain a contribution from spin which must be separated out as we will show. 

There is one point to notice about the two equations (\ref{eq:P}) and (\ref{eq:E}).  Although we are taking our algebras over the reals, we have introduced the {\em symbol} $i$.  We have deliberately called it a {\em symbol} because in the Schr\"{o}dinger case, ${\cal{C}}_{0,1}$, we use the generator $e$ in place of the symbol $i$, while in the Pauli case, ${\cal{C}}_{3,0}$ we use $e_{123}$.  We can do this because both elements are in the centre of their respective algebras.  The Dirac case is more complicated and that is one of the reasons  to discuss this specific case in a second paper \cite{bhbc081}.

\subsection{Time Evolution and the Generalized Dirac Derivatives.}

Let us now turn to examine the derivatives that can be introduced into the Clifford bundle we have constructed.  We can define two derivatives, one acting from the left and the other acting from the right.  These  so called generalised Dirac derivatives \cite{jgmm} take the form
\begin{eqnarray}
{\overrightarrow D}=\sum e_{i}\partial_{x_{i}}\hspace{0.5cm}\mbox{and}\hspace{0.5cm}{\overleftarrow D}=\sum \partial_{x_{i}}e_{i}.					\label{eq:conect}
\end{eqnarray}							
In these expressions, $e_{i}$ are the generators of the Clifford algebra, while $x_{i}$ are are the co-ordinates of the base manifold. Here the $e_{i}$ and the $x_{i}$ are related through the projection,  $\eta(e_{i})={\hat e}_{i}$,  where ${\hat e}_{i}$ are the set of orthonormal unit vectors that form the basis for the co-ordinate system on the base space \cite{bh08}.
In the non-relativistic case we will exclude  time from the sum in (\ref{eq:conect}) because in this case, time is treated, as usual, as an `external' parameter.  This special treatment will become unnecessary in relativistic theories.

Now we need to find time development equations for $\rho_{c} =\Phi_{L}\widetilde\Phi_{L}$. This means we  must consider equations involving derivatives of the form $({\overrightarrow D}\Phi_{L}){\widetilde\Phi_{L}}$ and $\Phi_{L}({\widetilde\Phi_{L}}{\overleftarrow D})$, where the $D$s, are defined in equation \eqref{eq:conect}.
Rather than treat these two derivatives separately, we will consider expressions like 
\begin{eqnarray*}
(e_{i}\;\partial_{x_{i}}\Phi_{L}){\widetilde\Phi_{L}}+\Phi_{L}(\partial_{x_{i}}{\widetilde\Phi_{L}}\;e_{i})
 \hspace{1cm}\mbox{and}\hspace{1cm}
(e_{i}\;\partial_{x_{i}}\Phi_{L}){\widetilde\Phi_{L}}-\Phi_{L}(\partial_{x_{i}}{\widetilde\Phi_{L}}\;e_{i})
\end{eqnarray*}
These derivatives do not include  time, so we also introduce  two  time derivatives 
\begin{eqnarray*}
(\partial_{t}\Phi_{L}){\widetilde\Phi_{L}}+\Phi_{L}(\partial_{t}{\widetilde\Phi_{L}})
\hspace{1cm}\mbox{and}\hspace{1cm}
(\partial_{t}\Phi_{L}){\widetilde\Phi_{L}}-\Phi_{L}(\partial_{t}{\widetilde\Phi_{L}})
\end{eqnarray*}

Now the dynamics must include the Dirac derivatives, the external potentials and the mass of the particle, so we will introduce two forms of the Hamiltonian, $\overrightarrow{H}=\overrightarrow{H}(\overrightarrow{D}, V, m)$ and $\overleftarrow{H}=\overleftarrow{H}(\overleftarrow{D}, V, m)$.
Our defining dynamical equations now read
\begin{eqnarray}
i[(\partial_{t}\Phi_{L}){\widetilde\Phi_{L}}+\Phi_{L}(\partial_{t}{\widetilde\Phi_{L}})]=i\partial_{t}\rho_{c}=(\overrightarrow{H}\Phi_{L}){\widetilde\Phi_{L}}-\Phi_{L}({\widetilde\Phi_{L}}\overleftarrow{H})
				\label{eq:conprob}
\end{eqnarray}							
and
\begin{eqnarray}
i[(\partial_{t}\Phi_{L}){\widetilde\Phi_{L}}-\Phi_{L}(\partial_{t}{\widetilde\Phi_{L}})]=(\overrightarrow{H}\Phi_{L}){\widetilde\Phi_{L}}+\Phi_{L}({\widetilde\Phi_{L}}\overleftarrow{H})
					\label{eq:anticom}
\end{eqnarray}							
Again we have introduced the {\em symbol} $i$ which has the meaning we discussed earlier.
The equations \eqref{eq:conprob} and \eqref{eq:anticom} can be written in the more compact form by writing
\begin{eqnarray*}
[H,\rho_{c}]_{\pm}=(\overrightarrow{H}\Phi_{L}){\widetilde\Phi_{L}}\pm\Phi_{L}({\widetilde\Phi_{L}}\overleftarrow{H})
\end{eqnarray*}
Then equation \eqref{eq:conprob} becomes
\begin{eqnarray}
i\partial_{t}\rho_{c}=[H,\rho_{c}]_{-}
				\label{eq:conprob2}		
\end{eqnarray}
In the Schr\"{o}dinger case this gives one equation which is simply the Liouville equation describing the conservation of probability.  In the Pauli case, this produces two equations, one is the conservation of probability equation, while the other conserves the total spin as the components develop in time as we will explain later.

Equation \eqref{eq:anticom} can be written in the form
\begin{eqnarray}
i\Phi_{L}\overleftrightarrow\partial_{t}{\widetilde\Phi_{L}}=[H,\rho_{c}]_{+}
					\label{eq:anticom2}		
\end{eqnarray}
Using equation (\ref{eq:E}), we find
\begin{eqnarray}
\rho E(t) =\alpha [H,\rho_{c}]_{+}	
					\label{eq:EH}  			
\end{eqnarray}

As far as we are aware equation (\ref{eq:anticom2}) has not appeared in the literature before it was introduced by Brown and Hiley \cite{mbbh}.  They arrived at this equation by a method different from the one used here and showed it was in fact the quantum Hamilton-Jacobi equation that appears in Bohm's approach to quantum mechanics. 

To summarise then, a complete specification of the dynamics is contained in these two equations,
the generalized Liouville equation (\ref{eq:conprob2}) 
and the quantum Hamilton-Jacobi equation (\ref{eq:EH}).
 Let us now see what these general equations give us in the Schr\"{o}dinger and Pauli cases.


\section{The  Schr\"{o}dinger particle.}


\subsection{The Preliminaries}


Let us first consider the Schr\"{o}dinger particle.  We start with the Clifford algebra, ${\cal C}_{0,1}$, taken over the reals. The algebra is generated by the  elements $\{1,e\}$ where $e^{2}=-1$.  A general expression for  an element of a left ideal is given by $\Phi_{L}=\phi_{L}\epsilon$ where $\epsilon$ is a primitive idempotent.  There is only one idempotent in ${\cal{C}}_{(0,1)}$, namely, $\epsilon = 1$ which means 
 we can write $\phi_{L}=R(g_{0}+g_{1}e)=RU$, where $g_{0}(\boldsymbol r, t)$ and $g_{1}(\boldsymbol r, t)$ are scalar real functions.  $R(\boldsymbol r, t)$ is a real scalar.
 
We now introduce the Clifford conjugate to $\Phi_{L}$, namely, $\Phi_{R} = {\widetilde \Phi_{L}}$.  This means that $\widetilde\Phi_{L}={\widetilde U}{\widetilde R}=R{\widetilde U}$ where  $\widetilde{U}=g_{0}-g_{1}e$.  Combining these results we find that $\rho_{c}=\Phi_{L}{\widetilde\Phi_{L}}={\phi_{L}}\widetilde\phi_{L}=R^2$, so that $U\widetilde{U}=\widetilde{U}U=1$. 

We can now relate these results to the standard approach by using the relations 
\begin{eqnarray*}
2g_{0}=\psi +\psi^{*}\hspace{1cm}\mbox{and}\hspace{1cm}
2eg_{1}=\psi - \psi^{*}.
\end{eqnarray*}
Here $\psi $ is the ordinary wave function.  This means that
\begin{eqnarray*}
\rho_{c}=\psi^{*}\psi=R^{2} =\rho
\end{eqnarray*}
Thus in the case of the Schr\"{o}dinger particle, the Clifford density element is simply the probability.  At first sight it seems we have gained no advantage over the conventional approach.  However in the case for the Pauli and Dirac particles, we find an essential difference between the Clifford density element and the probability.

Notice in this approach we have replaced the symbol  $i$ by $e$.  As we have remarked before it is this replacement that enables us to embed the Schr\"{o}dinger formalism in the algebra ${\cal C}_{0,1}$ taken over the {\em reals}.
The reason why this works is because the real Clifford algebra ${\cal C}_{0,1}$ is isomorphic to the complex numbers.

Returning to the Clifford density element we see it  becomes
\begin{eqnarray}
\rho_{c}=\Phi_{L}{\widetilde\Phi_{L}}=\phi_{L}\epsilon{\widetilde\phi_{L}}\label{eq:bl1}							
\end{eqnarray}
Since $\epsilon=1$,  the density operator becomes the usual expression for the probability density of a pure state.

The expression (\ref{eq:bl1}) gives us the only bilinear invariant of the first kind that can be formed in  ${\cal C}_{0,1}$.  Clearly this is not sufficient to completely specify the state of the quantum process.  As we explained in section 2.4 we need to construct a bilinear invariant of the second kind. This we do in the next sub-section.

\subsection{The Energy and Momentum Densities}


To construct bilinear invariants of the second kind we will use equations (\ref{eq:P}) and (\ref{eq:E}) with $\alpha =1/2$.  We can then write  
the components of the momentum, $P^{j}$ as
\begin{eqnarray}
2\rho P^{j}(t)= -e[\left(\partial^{j}(RU)\right)\epsilon R\widetilde{U}-RU\epsilon\partial^{j}(R\widetilde{U})]						
\end{eqnarray}
This  can then be reduced to
\begin{eqnarray}
2P^{j}(t)=-e\Omega^{j}\cdot\Sigma 			
				\label{eq:P3}		
\end{eqnarray}
where $\Sigma = U\epsilon \widetilde{U}$ with $\epsilon =1$.  In this equation we have written the scalar product as
\begin{eqnarray}
\Omega\cdot\Sigma=\frac{1}{2}[\Omega\Sigma+\Sigma\Omega]\nonumber
\end{eqnarray}
Here we have followed Hestenes and Gurtler \cite{dh71} and defined
\begin{eqnarray}
\Omega^{j}=2(\partial^{j}U)\widetilde{U}=-2U(\partial^{j}\widetilde{U}).
				\label{eq:WJ}			
\end{eqnarray}
In the Schr\"{o}dinger case, we have $\Sigma =1$ so that
\begin{eqnarray}
2P^{j}(t)=-e\Omega^{j}					
\end{eqnarray}
If we write $U=\exp(eS)$ so that  $g_{0}=\cos S$ and $g_{1}=\sin S$, then we find
\begin{eqnarray}
P^{j}(t)=\partial^{j}S\hspace{1cm}\mbox{or}\hspace{1cm} {\bf P} (t)={ \boldsymbol\nabla}  S
\end{eqnarray}
Thus, as we have remarked earlier,  $ {\bf P}$  gives an expression which we can identify with the Bohm momentum.

We can go through a similar procedure for the energy.  It is easy to show
\begin{eqnarray}
2E(t)=e[\Omega_{t}\Sigma+\Sigma\Omega_{t}]	
				\label{eq:E3}			
\end{eqnarray}
where $\Omega_{t}=2(\partial_{t}U){\widetilde U}=-2U(\partial_{t}{\widetilde U})$.
As before writing  $\epsilon=1$ and $U=\exp[eS]$ we find
\begin{eqnarray}
E(t)=-\partial_{t}S.			\label{eq:EB}		
\end{eqnarray}
which is, again, the expression for the energy used in the standard Bohm approach.

\subsection{The Time Evolution of the Schr\"{o}dinger Particle.}


Now we can go further and use equation (\ref{eq:EB}) in equation (\ref{eq:EH}), again with $\alpha =1/2$, to obtain the quantum Hamilton-Jacobi equation 
\begin{eqnarray}
-2\partial_{t}S=[H,\rho_{c}]_{+}				
\end{eqnarray}
If we use the Hamiltonian $H=p^{2}/2m+V$ and use the $p=\nabla S$ we find
\begin{eqnarray}
\partial_{t}S+(\nabla S)^{2}/2m+Q+V=0			
\end{eqnarray}
where $Q=-\nabla^{2}R/2mR$ which is immediately recognised as the quantum potential.

Together with equation (\ref{eq:conprob2}), which we write in the form   
\begin{eqnarray}
e\partial_{t}\rho_{c}+[\rho_{c},H]_{-}=0			
\end{eqnarray}
This is immediately recognised as the Liouville equation which shows that the probability is conserved as required.


\subsection{The Current}


Let us introduce the current, $J$, defined by
\begin{eqnarray}
mJ_{i}=(\partial_{i}\Phi_{L}){\widetilde\Phi_{L}}+\Phi_{L}(\partial_{i}{\widetilde\Phi_{L}})
\end{eqnarray}							
which reduces
\begin{eqnarray}
mJ_{i}=-e\rho[\Omega_{i}\Sigma +\Sigma\Omega_{i}]
\end{eqnarray}							
If we  put $\epsilon=1$, so that $\Sigma=1$, then $J_{i}=-e\rho\Omega_{i}/2m$ and if we again write $U=\exp(eS)$, we find $J_{i}=\rho\partial_{i}S/m$.  But the velocity of the particle is defined through
\begin{eqnarray}
mv_{i}=mJ_{i}/\rho						
\end{eqnarray}
which means that
\begin{eqnarray}
mv=\nabla S.			\label{eq:mv}		
\end{eqnarray}
We can then define the trajectories in the phase space we have constructed by integrating equation (\ref{eq:mv}).  The form of these trajectories are well known \cite{pdh}.  In this way we complete the results for the Schr\"{o}dinger theory.

 We see that all the results of  conventional quantum mechanics and of the Bohm model can be obtained directly from the Clifford algebra.  At this stage it looks as if we have replaced one formalism by a more complicated formalism without gaining any benefit. However we will now show how the Pauli particle can be treated in a similar way, only now using a larger Clifford algebra ${\cal C}_{3,0}$.  This then opens the possibility of including the Dirac particle in such an analysis but now using a yet larger Clifford algebra ${\cal C}_{1,3}$ as we will show in a later paper \cite{bhbc081}.


\section{The Pauli Particle.}


\subsection{The Clifford Density Element.}


In the Pauli case, we use the Clifford algebra generated by the elements $\{1, e_{1},e_{2},e_{3}\}$ with the multiplication rule $e_{i}e_{j}+e_{j}e_{i}=2\delta_{ij}$.  This forms the Pauli Clifford, ${\cal C}_{3,0}$\footnote{This is also written as C(2) in Porteous\cite{ip}.}. In this algebra, an element of the left ideal again takes the form $\Phi_{L}=\phi_{L}\epsilon$ and its Clifford conjugate can be written in the form $\widetilde\Phi_{L}= \tilde\epsilon{\widetilde \phi}_{L} $.  Here $\epsilon$ is some primitive idempotent in the algebra.  The idempotents that we will use satisfy the condition $\epsilon=\tilde\epsilon$.

The choice of idempotent is determined by the physical context as we will show below.   Although $\phi_{L}$  can be any element of the Clifford algebra, it can be shown that it is sufficient to restrict  $\phi_{L}$  to any  even element of the Clifford algebra\footnote{By even we mean any element that can be expanded in terms of an even number of basis elements including the scalar.}.  We will again write $\phi_{L}=RU$ but here 
\begin{eqnarray}
U=g_{0}+g_{1}e_{23}+g_{2}e_{13}+g_{3}e_{12}\hspace{1cm}\mbox{and}\hspace{1cm}\sum^{3}_{i=0}g^{2}_{i}=1. 
				\label{eq:U-g}		
\end{eqnarray}							
[Note here $e_{ij}\equiv e_{i}e_{j}$.]  In this case $\widetilde{U}=g_{0}-g_{1}e_{23}-g_{2}e_{13}-g_{3}e_{12}$ so that again we have $U\widetilde{U}=\widetilde{U}U=1$.
Since the CDE is $\rho_{c}=\Phi_{L}{\widetilde\Phi_{L}}=\phi_{L}\epsilon{\widetilde\phi_{L}}$, it now takes the simple form 
$
\rho_{c}=\rho U\epsilon \widetilde{U}$, where
$\rho =R^{2}$ is again  the probability density.

To proceed further we need to chose the form of the idempotent $\epsilon$.  To do this we need to first choose a particular direction in space.  This of course can be any direction in space, but usually this choice will be determined physically by introducing a suitable uniform magnetic field. We take this  direction to be the 3-axis, even in the absence of any field.  Consequently we choose the idempotent to be $\epsilon =(1+e_{3})/2$.    

With this choice the CDE becomes
\begin{eqnarray}
\rho_{c}=\Phi_{L}{\widetilde\Phi_{L}}=\phi_{L}\epsilon{\widetilde\Phi_{L}}=\rho U\epsilon\tilde{U}=\rho(1+Ue_{3}\widetilde{U})/2		
\end{eqnarray}
where $Ue_{3}\widetilde{U}$ is a vector which turns out to be the spin vector ${\bf s} = (a_{1}e_{1}+a_{2}e_{2}+a_{3}e_{3})/2$.  

To show this, let us first choose an element of the algebra  defined by
\begin{eqnarray}
\hat{S}=\Phi_{L}e_{12}{\widetilde\Phi_{L}}= \phi_{L}\epsilon e_{12}\epsilon{\widetilde\Phi_{L}}=\rho(U\epsilon e_{12}\epsilon\widetilde{U})		
\end{eqnarray}
But $\epsilon e_{12}\epsilon=e_{12}\epsilon$, so that 
\begin{eqnarray}
\hat{S}=i\rho(1+Ue_{3}\widetilde{U})/2				
\end{eqnarray}
We will call  the vector part of $\hat{S}$, the spin bivector, namely,  ${\cal S}=iUe_{3}\widetilde{U}$ and we will call 
\begin{eqnarray}
{\bf s}=(Ue_{3}\widetilde{U})/2					
\end{eqnarray}
the spin.  Each component of the spin is given by the Clifford scalar  
\begin{eqnarray}
e_{i}\cdot{\bf s}=(e_{i}Ue_{3}\widetilde{U}+Ue_{3}\widetilde{U}e_{i})/2.
\end{eqnarray}							
This means that $e_{i}\cdot {\bf s}=a_{i}/2$ where the $a_{i}$ are given by
\begin{eqnarray}
a_{1}=2(g_{1}g_{3}-g_{0}g_{2})\hspace{0.5cm}a_{2}=2(g_{0}g_{1}+g_{2}g_{3})\hspace{0.5cm}a_{3}=g_{0}^{2}-g_{1}^{2}-g_{2}^{2}+g_{3}^{2}.					
				\label{eq:a-g}			
\end{eqnarray}

To show this is exactly the same object as used in the conventional approach, let us convert this into a matrix representation.  In the usual representation, the spinor is written in the form
\begin{eqnarray}
\Psi=\begin{pmatrix}
     \psi_{1}   \\
      \psi_{2}
\end{pmatrix}
				\label{eq:matspin}
\end{eqnarray}							
Here $\psi_{1}$ and $\psi_{2}$ are complex numbers.  Again in this representation, the $\{e_{i}\}$ are represented by the usual Pauli spin matrices, $\{\sigma_{i}\}$ and we find that the set $\{g_{i}\}$ are related to the $\psi s$ by
\begin{eqnarray}
g_{0}= (\psi_{1}^{*}+\psi_{1})/2\hspace{0.5cm}g_{1}=i(\psi_{2}^{*}-\psi_{2})/2 \nonumber\\ g_{2}=(\psi_{2}^{*}+\psi_{2})/2
\hspace{0.5cm}g_{3}=i(\psi_{1}^{*}-\psi_{1})/2	
				\label{eq:g-psi}			
\end{eqnarray}
Using these relations it is not difficult show that when we write $s=\sum_{j}a_{j}e_{j}$ the coefficients are
\begin{eqnarray}
a_{1}=\psi_{1}\psi^{*}_{2}+\psi_{2}\psi^{*}_{1}\hspace{0.75cm}
a_{2}=i(\psi_{1}\psi^{*}_{2}-\psi_{2}\psi^{*}_{1})\hspace{0.75cm}
a_{3}=|\psi_{1}|^{2}-|\psi_{2}|^{2}			
\end{eqnarray}
This is just the usual well known expression for the spin vector when written in matrix form.  

Now we can write the CDE in the form
\begin{eqnarray}
\rho_{c}=\rho(1+{\bf s.e})/2				
\end{eqnarray}
When normalised with $\rho=1$, and the $\{e_{i}\}$ replaced by the Pauli matrices $\{\sigma_{i}\}$, this operator becomes the standard expression for the density matrix  \cite{lb}.  Thus up to now we are simply writing  the conventional formalism in a general representation in which the choice of a specific matrix representation is not necessary.

\subsection{The Bohm Momentum and Energy for the Pauli Particle.}


In the usual discussion of the Pauli theory, we generally assume the particle is charged and coupled to the electromagnetic field through the vector potential.  In order to keep the formalism as simple as possible so that we can bring out clearly the quantum aspects of our approach, we examine the behaviour of the particle in the absence of an electromagnetic field.  Once the principles involved in our approach have been brought out clearly, it is easy then to introduce the electromagnetic coupling through the minimal coupling $\nabla \rightarrow \nabla -eA$.

We start by considering the momentum, $ P^{j}$, defined by equation  (\ref{eq:P}) with $\alpha=1$.  Thus  substituting  $ \rho_{c}=\phi_{L}\epsilon{\widetilde\phi_{L}}$ into this equation, we find 
 \begin{eqnarray}
P^{j}(t)=-\frac{i}{2}[\Omega^{j}\Sigma +\Sigma \Omega^{j}]=
-i\Omega^{j}\cdot\Sigma	   				
					\label{eq:P4}		
\end{eqnarray}
Equation (\ref{eq:E}), again with $\alpha=1$, becomes
\begin{eqnarray}
E(t) = \frac{i}{2}[\Omega_{t}\Sigma+\Sigma\Omega_{t}]=i\Omega_{t}\cdot \Sigma.	\label{eq:P4a}												
\end{eqnarray}
Here we have again defined $\Sigma = U\epsilon {\hat U}$ and $\Omega^{j}=2(\partial ^{j}U)\tilde{U}$.  The equations \eqref{eq:P4} and \eqref{eq:P4a}  should be compared with the corresponding equations for the Schr\"{o}dinger case (\ref{eq:P3}) and (\ref{eq:E3}) .  The only difference lies in the choice of $\epsilon$.  

In the Pauli case we choose $\epsilon = (1+e_{3})/2$.  Substituting this into equation (\ref{eq:P4}) gives
\begin{eqnarray}
P^{j}(t)=-\Omega^{j}\cdot S-i\Omega^{j}/2		 
\end{eqnarray}
where $S=i(Ue_{3}\tilde{U})/2$ is the spin bivector.  In the Pauli case, we write $i=e_{123}$ since it commutes with all the elements of the real Pauli algebra and $(e_{123})^{2}=-1$.  The Bohm momentum is the scalar part of this expression so that
\begin{eqnarray}
 P_{B}(t)=-\Omega\cdot  S			
 					\label{eq:P5}		
\end{eqnarray}
where $\Omega=\sum \Omega^{j}$.
Equation (\ref{eq:P5}) is exactly the equation (2.11) of Hestenes and Gurtler \cite{dh71}.
The corresponding scalar part of the energy becomes
\begin{eqnarray}
E_{B}(t)=\Omega_{t}\cdot  S				
					\label{eq:E5}		
\end{eqnarray}
which corresponds to equation (2.6) of Hestenes and Gurtler \cite{dh71}.

We must now use equation (\ref{eq:U-g}) together with (\ref{eq:a-g}) and (\ref{eq:g-psi}) to show that
\begin{eqnarray}
2\rho  P_{B}(t)=i[(\nabla\psi_{1})\psi^{*}_{1}-(\nabla\psi^{*}_{1})\psi_{1}+(\nabla\psi_{2})\psi^{*}_{2}-(\nabla\psi^{*}_{2})\psi_{2}]						         			
\end{eqnarray}
Writing $\psi_{1}=R_{1}e^{iS_{1}}$ and $\psi_{2}=R_{2}e^{iS_{2}}$, equation \eqref{eq:P4}  becomes
\begin{eqnarray}
\rho  P_{B}(t)=(\nabla S_{1})\rho_{1}+(\nabla S_{2})\rho_{2}
\end{eqnarray}							
where $\rho_{i}=R_{i}^{2}$. The meaning becomes more transparent if we write $ P_{i}=\nabla S_{i}$ when the expression for the momentum becomes
\begin{eqnarray}
\rho  P_{B}(t)= P_{1}\rho_{1} +  P_{2}\rho_{2}		
				\label{eq:MP}			
\end{eqnarray}
Thus we see that in terms of the usual approach $P_{B}(t)$ is the weighted mean of the momentum that can be attributed to each component of the spinor acting by itself.  This result was already noted in Bohm and Hiley \cite{bh93}.

Similarly the energy becomes
\begin{eqnarray}
\rho E_{B}(t)=-[(\partial_{t}S_{1})\rho_{1}+(\partial_{t}S_{2})\rho_{2}]
\end{eqnarray}							
Using the polar form for each spinor component $\psi_{i}=R_{i}e^{iS_{i}}$, the energy can be written in the form
\begin{eqnarray}
\rho E_{B}=E_{1}\rho_{1}+E_{2}\rho_{2}	
				\label{eq:ME}		
\end{eqnarray}
which is clearly seen as the weighted mean of the energy associated with each component of the spinor.

We have remaining the vector part for both $P$ and $E$ defined by equations (\ref{eq:P}) and (\ref{eq:E}).  We see from the earlier work that $\Omega^{j}=2(\partial^{j}U){\widetilde U}=-2U(\partial^{j}{\widetilde U})$ and $\Omega_{t}=2(\partial_{t}U){\widetilde U}=-2U(\partial_{t}{\widetilde U})$ which implies that $\Omega$ appears to be a form of angular velocity.  Indeed if we express the components $\psi_{1}$ and $\psi_{2}$ in terms of Euler angles as explained in the next section, we find $\Omega$ is exactly the expression for the angular velocity of a rotating frame.  This result suggests that we can describe the spinning electron in terms of Cartan's moving frames \cite{hf}, a feature that Hestenes \cite{dh03} exploits.


\subsection{Relationship to the Bohm-Schiller-Tiomno model}


As we remarked in section 2, Bohm, Schiller and Tiomno,  [BST], \cite{BST} 
obtained expressions for the Bohm momentum and energy using a generalisation of the polar decomposition of the wave function as used in the Schr\"{o}dinger case.  This was possible here because the spinor, when written in terms of Euler angles, takes the form
\begin{eqnarray}\Psi=\begin{pmatrix}
      \psi_{1}    \\
      \psi_{2}  
\end{pmatrix}=
\begin{pmatrix}
     \cos(\theta/2) \exp(i\phi/2)    \\
      i\sin(\theta/2) \exp(-i\phi/2)
\end{pmatrix}\exp(i\psi/2)
					\label{eq:matspin2}	
\end{eqnarray}
Then BST write this spinor in the form $\Psi=\Phi e^{iS}$, thus implicitly identifying the Euler angle $\psi$ with an overall phase $S$.  We have always felt the meaning of this step was unclear.  Using this assumption, they then found that the Bohm momentum can be written in the form
\begin{eqnarray}
 P_{B}=(\nabla S+\cos\theta\nabla\phi)/2		
 					\label{eq:P6}		
\end{eqnarray}
while the energy was written in the form
\begin{eqnarray}
E_{B}(t)=	-(\partial_{t} S +\cos \theta\partial_{t}\phi)/2
					\label{eq:E6}	
\end{eqnarray}
Both of these expressions reduce to usual the Schr\"{o}dinger result when $\theta = \phi = 0. $ 

These expressions for the Bohm momentum and energy can be shown to be identical to the corresponding two expressions (\ref{eq:P5}) and (\ref{eq:E5}).  This identity can be  confirmed by a tedious but straight forward method by converting the spinor shown in equation (\ref{eq:matspin}) into the form shown in equation (\ref{eq:matspin2}). In this way we see that there is no need to single out the Euler angle $\psi$ and treat it as a common phase.  The method used to establish (\ref{eq:P5}) and (\ref{eq:E5}) is thus more general in the sense that we do not have to appeal to a specific representation in terms of Euler angles.  We will show in the next paper that this is the clue that will enable us to extend the Bohm approach to the Dirac equation.


\section{The Pauli Hamilton-Jacobi equation.}

To obtain an expression for the Pauli quantum Hamilton-Jacobi equation, we need to examine equation (\ref{eq:EH}) with $\rho_{c}=\phi_{L}\epsilon{\widetilde\phi_{L}}$ where now $\epsilon = (1+e_{3})/2$.  In this case $\Phi_{L}\overleftrightarrow\partial_{t}\widetilde\Phi_{L}$ splits into two parts, a scalar part and a vector part.  The scalar part is
\begin{eqnarray}
2\langle\Phi_{L}\overleftrightarrow\partial_{t}\widetilde\Phi_{L}\rangle_{s}= (\partial_{t}\phi_{L})e_{12}{\widetilde\phi_{L}}- \phi_{L}e_{12}(\partial_{t}\widetilde\phi_{L})
\end{eqnarray}						
The vector part is
\begin{eqnarray}
2\langle\Phi_{L}\overleftrightarrow\partial_{t}\widetilde\Phi_{L}\rangle_{v}=(\partial_{t}\phi_{L})e_{123}{\widetilde\phi_{L}}-\phi_{L}e_{123}(\partial_{t}{\widetilde\phi_{L}})
\end{eqnarray}						
Similarly we can split the RHS of equation (\ref{eq:EH}) into a scalar part
\begin{eqnarray}
2\langle[H,\rho_{c}]\rangle_{s}=(H\phi_{L}){\widetilde\phi_{L}}+\phi_{L}(H{\widetilde\phi_{L}})
\end{eqnarray}						
while the vector part is
\begin{eqnarray}
2\langle[H,\rho_{c}]\rangle_{v}=(H\phi_{L})e_{3}{\widetilde\phi_{L}}+\phi_{L}e_{3}(H{\widetilde\phi_{L}})
\end{eqnarray}						
This separation appears to  produce two independent equations.  However on closer examination, we find their content is identical, so we need to investigate only one of them, the scalar equation
\begin{eqnarray}
(\partial_{t}\phi_{L})e_{12}{\widetilde\phi_{L}}- \phi_{L}e_{12}(\partial_{t}{\widetilde\phi_{L}})=(H\phi_{L}){\widetilde\phi_{L}}+\phi_{L}(H{\widetilde\phi_{L}})
\end{eqnarray}						
However we have already evaluated the LHS of this equation in working out the Bohm energy, namely
\begin{eqnarray}
\rho E_{B}(t)=(\partial_{t}\phi_{L})e_{12}{\widetilde\phi_{L}}- \phi_{L}e_{12}(\partial_{t}{\widetilde\phi_{L}})=\rho \Omega_{t} \cdot S.
\end{eqnarray}						
To evaluate the RHS we assume a free particle Hamiltonian\footnote{The inclusion of an interaction with the electromagnetic field is straight forward.  However we again omit the details in this presentation to keep the everything as simple as possible.} $H=-\nabla^{2}/2m$.  We find
\begin{eqnarray}
2m[(H\phi_{L}){\widetilde\phi_{L}}+\phi_{L}(H{\widetilde\phi_{L}})]=\rho[S\wedge \nabla P+S\cdot \nabla  W + P\cdot P +  W\cdot  W]
\end{eqnarray}						
where we have introduced the shorthand $A\wedge B=(AB-BA)/2$.  Here $W=\rho^{-1}\nabla(\rho S)$.  It is not difficult to show that $S\wedge \nabla=0$ so that we end up with the equation
\begin{eqnarray}
2mE(t)=P^{2}+[2( \nabla W\cdot S) + W^{2}]
\end{eqnarray}						
This is the Pauli quantum Hamilton-Jacobi equation where the quantum potential is\begin{eqnarray}
Q= (\nabla W\cdot S)/m + W^{2}/2m
						\label{eq:QP}
\end{eqnarray}							
Now let us express equation (\ref{eq:QP}) purely in terms of $ P, \rho$ and the spin bivector $S$.  After some straight forward but tedious 
work we find 
\begin{eqnarray}
Q= \{S^{2}[2\nabla^{2}\ln\rho +( \nabla\ln\rho)^{2}] +S\cdot\nabla^{2}S\}/2m=Q_{1} +Q_{2}
						\label{eq:QPCD}
\end{eqnarray}							
An expression for $Q$ has been evaluated by Dewdney {\em et al.} \cite{dhk86} in terms of Euler angles and we will show that equation (\ref{eq:QPCD}) reduces to exactly the same expression that they find.  To show this we again need to use the relations that convert equation (\ref{eq:matspin}) into the Euler angle representation (\ref{eq:matspin2}).  Then we can show that 
\begin{eqnarray}
Q_{1}=-\frac{1}{2m}\frac{\nabla^{2}R}{R}		
\end{eqnarray}
Here we recognise the quantum potential contribution to the Schr\"{o}dinger particle.  However we now have a spin dependent part, $Q_{2}$ which takes the form
\begin{eqnarray}
Q_{2}=[(\nabla\theta)^{2} + \sin^{2}\theta(\nabla\phi)^{2}]/8m
\end{eqnarray}
Thus we can write the Pauli quantum Hamilton-Jacobi equation in the form
\begin{eqnarray}
(\partial_{t}\psi +\cos \theta \partial_{t}\phi)/2 +P_{B}^{2}/2m +Q_{1} +Q_{2} =0								
\end{eqnarray}
which, in this representation, agrees exactly with the expression given in Dewdney {\em et al.} \cite{dhk86}.


\subsection{The Quantum Torque}

We start from the Pauli quantum Hamilton-Jacobi equation
\begin{eqnarray}
E_{B}(t)=P_{B}^{2}/2m + Q
\end{eqnarray}						
and differentiate to form
\begin{eqnarray}
\nabla E_{B}(t)=\nabla (P_{B}^{2})/2m +\nabla Q
				\label{eq:gradE}
\end{eqnarray}						
Now we also have
\begin{eqnarray}
\partial_{t} P_{B} +\nabla E_{B}=e_{k}\Omega_{t}\cdot\partial_{k}S
\end{eqnarray}						
Rather than working with the general expression which can be found in Hestenes and Gurtler \cite{dh71}, it is simpler to work from the specific forms of $ P_{B}(t)$ and $E_{B}(t)$ given in equations (\ref{eq:P6}) and (\ref{eq:E6})  then we find
\begin{eqnarray}
\partial_{t}P_{B}+\nabla E_{B}=\frac{1}{2}[\partial_{t}(\cos\theta)\nabla\phi - \nabla(\cos\theta)\partial_{t}\phi]			
\end{eqnarray}
Using this in equation (\ref{eq:gradE}) we find
\begin{eqnarray}
\frac{dP_{B}}{dt}=-\nabla Q-\frac{1}{2}[\partial_{t}(\cos\theta)\nabla\phi - \nabla(\cos\theta)\partial_{t}\phi]			
\end{eqnarray}
Thus we see that the `quantum force' does not come simply from the gradient of the quantum potential as in the case of the Schr\"{o}dinger particle, but has an additional intrinsic `torque' coming from the rotational motion as specified by the Euler angles $\theta$ and $\phi$.  If need be, we could keep the whole expression  general and use $R_{1}, R_{2}, S_{1}$ and $S_{2}$.  However we have no need for these equations in this paper so we will not present them here.


\subsection{The Conservation Equations for the Pauli Particle.}

So far we have only investigated the dynamical consequences of equation (\ref{eq:EH}).  We must now turn to equation (\ref{eq:conprob2}) which we have called the generalised Liouville equation.  As in the case of the quantum Hamilton-Jacobi equation (\ref{eq:EH}), equation (\ref{eq:conprob2}) contains two equations, again corresponding to a scalar part  and a bivector part.  We will now show that the scalar equation gives us a conservation of probability equation, a genuine Liouville equation.  The other gives us an equation involving the spin of the particle. 

We find the LHS of equation (\ref{eq:EH}) becomes
\begin{eqnarray}
i[\partial_{t}\rho_{c}=i\partial_{t}[\rho+\phi_{L}e_{3}{\widetilde\phi_{L}}]=
i\partial_{t}\rho+2\partial_{t}(\rho S)
\end{eqnarray}							
where $\rho$ is the usual probability.  In this expression, the term $i\partial_{t}\rho$ corresponds to the scalar part, while $\partial_{t}(\rho S)$ is the bivector part.

The RHS of equation (\ref{eq:conprob2}), $[H,\rho_{c}]_{-}$ also splits into a scalar and a bivector part, the scalar part being
\begin{eqnarray}
\langle[H,\rho_{c}]_{-}\rangle_{s}=(H\phi_{L})e_{3}{\widetilde\phi_{L}}-
\phi_{L}e_{3}(H{\widetilde\phi_{L}})
\end{eqnarray}							
We will again, for simplicity, use the free particle Hamiltonian and after some manipulation we find
\begin{eqnarray}
2m\langle[H,\rho_{c}]_{-}\rangle_{s}=-i\rho\{2\nabla P-[S(P\cdot W)+(P\cdot W)S]\}
\end{eqnarray}							
Using the expression for $W=\rho^{-1}\nabla(\rho S)$ given above and after some work, we find
\begin{eqnarray}
\partial_{t}\rho + \nabla(\rho P/m)=0
\end{eqnarray}							
This will immediately be recognised as the Liouville equation for the conservation of probability.

Now let us turn to the bivector part of the equation (\ref{eq:EH}). We need to combine $\partial_{t}(\rho S)$ with the bivector part of $[H,\rho_{c}]_{-}$. We find
\begin{eqnarray}
4m\partial_{t}(\rho S)=2m\langle[H,\rho_{c}]_{-}\rangle_{B}=4\rho[(\nabla P\cdot S)+(S\wedge \nabla W)+(P\cdot W)]
\end{eqnarray}							
Since 
\begin{eqnarray}
2P\cdot W=(\nabla\ln\rho)P\cdot S +2(P\cdot\nabla)S
\end{eqnarray}							
we find
\begin{eqnarray}
\left[\partial_{t} +\frac{P\cdot \nabla}{m}\right]S=2(\nabla\wedge S)
					\label{eq:dpdt}
\end{eqnarray}							
However simplifying the RHS of equation (\ref{eq:dpdt}) we find
\begin{eqnarray}
\nabla W\wedge S=(\nabla\ln\rho)(\nabla S\wedge S) +\nabla^{2}S\wedge S
\end{eqnarray}							
so that finally we get
\begin{eqnarray}
\left[\partial_{t} +\frac{P\cdot \nabla}{m}\right]S=\frac{dS}{dt}=\frac{1}{m}
\left[ \nabla^{2}S +(\nabla\ln\rho)\nabla S\right]\wedge S
				\label{eq:dpdt2}			
\end{eqnarray}
To connect up with the work of Dewdney {\em et al.} we now write this equation in terms of the spin vector $s$ rather than the spin bivector $S$.  We can do this because $S=is$ so that equation (\ref{eq:dpdt2}) becomes
\begin{eqnarray}
\frac{ds}{dt}=\frac{s}{m}\times[\nabla^{2}s +(\nabla\ln\rho)\nabla s].				\label{eq:SC}			
\end{eqnarray}
where we have used the identity $A\wedge B=i(A\times B)$.  This shows that a quantum torque acting on the components of the spin.  Equation (\ref{eq:SC}) then ensures the total spin is conserved during the time development.  This equation was exploited numerically by Dewdney {\em et al} \cite{dhk86} to show how the spin turned as it passed through an inhomogenious magnetic field.


\subsection{The Pauli Current.}

We will complete our discussion of the Pauli particle by deriving an expression for the current.  We begin by using the definition of the current found in Messiah \cite{am}.  In our notation this becomes
\begin{eqnarray}
mJ_{k}=\sum_{j}[e_{k}(e_{j}p_{j}\Phi_{L})\epsilon{\widetilde\phi_{L}}+\Phi_{L}\epsilon e_{j}e_{k}(p^{\dag}_{j}{\widetilde\phi_{L}})]		
\end{eqnarray}
The expression for $J_{k}$ splits into two parts, a convectional part, $ J_{\mbox{conv}}$ and a rotational part, $ J_{\mbox{rot}}$.  We find
\begin{eqnarray}
 mJ_{\mbox{conv}}=-i\rho\Omega\cdot\Sigma		
\end{eqnarray}
Once again if we take the scalar part, we find 
\begin{eqnarray}
 mJ_{\mbox{conv}}=-\rho(\Omega\cdot S)=\rho  P_{B}(t)
\end{eqnarray}
where we have used equation (\ref{eq:P5}) to identify the convection current with the Bohm momentum.

Turning to the rotational part of the current, we find
\begin{eqnarray}
mJ_{\mbox{rot}}=-i\sum_{j}[e_{kj}(\partial_{j}\phi_{L})\epsilon{\widetilde\phi_{L}}-\phi_{L}\epsilon e_{jk}(\partial_{j}{\widetilde\phi_{L}})]=i\sum_{j}e_{jk}\partial_{j}[\phi_{L}\epsilon{\widetilde\phi_{L}}]			
\end{eqnarray}
so that
\begin{eqnarray}
mJ_{\mbox{rot}}=-i\nabla\wedge(\rho\Sigma)=\nabla\times(\rho\Sigma)								
\end{eqnarray}
Now we note $\Sigma =1/2+S$ so that the scalar part of the rotational part of the current is
\begin{eqnarray}
mJ_{\mbox{rot}}=\nabla\times(\rho S)			
\end{eqnarray}
If we write $J=\rho v$ we obtain
\begin{eqnarray}
m\rho v=\rho P_{B} +\nabla\times(\rho S)				
\end{eqnarray} 
Thus we see the momentum of the particle comprises two parts corresponding to a linear motion and a rotational part.



\section{Conclusions}

In this paper we have shown the results of standard quantum mechanics for the Schr\"{o}dinger and Pauli theories can be derived from Clifford algebras without the need to introduce the notion of a wave function.   All the information normally contained in the wave function can be encoded in an element of a minimal left ideal, $\Phi_{L}$ of the appropriate Clifford algebra. 
To use this information in an invariant way, we construct what we have called a {\em Clifford density element} [CDE] by forming the product $\rho_{c} = \Phi_{L}\Phi_{R}$, where $\Phi_{R}=\widetilde{\Phi_{L}}$ is an element of the conjugate minimal right ideal defined by Clifford conjugation.  We then  introduce two equations, (\ref{eq:conprob2}) and (\ref{eq:anticom2}) that describe the time evolution of $\rho_{c}$. These two equations describe explicitly the time development of the phase and amplitude information appearing in the conventional approach.

We then show how the bilinear invariants of the first and second kind used in the usual approach are encoded in the algebra and use these invariants to describe the physical process.  We illustrated this by forming the scalar quantities $P^{j}$ and $E$ given by the defining equations (\ref{eq:P}) and (\ref{eq:E}).  These equations can be derived from the standard expressions for the momentum density and the energy density and contain information about the Bohm momentum and the Bohm energy.  In the case of the Pauli theory equation (\ref{eq:P})  also contains information about the spin of the  particle.

In this way we have shown that all the usual results for the Schr\"{o}dinger and Pauli theories can be obtained from the structure of Clifford algebras.   This provides a unified mathematical structure without the need to refer to an external Hilbert space.  Nevertheless it is open to us to represent the results of our algebraic approach in such a space, but this is by no means necessary.
 
Furthermore we recover all the equations used in the Bohm models for the Schr\"{o}dinger and Pauli theory \cite{db52}, \cite{BST}, \cite{bh93}, \cite{ph93}. Not only do we obtain the essential equations, but we obtain them in a systematic way in which it is not necessary to start from the wave function. That is, it is not necessary to start from a polar decomposition of the wave function, a procedure that gives Bohm's original model the appearance of  a superficial and even artificial approach to quantum phenomena.  Instead we find that the Clifford algebra itself carries all the necessary information without ever mentioning the wave function.  

One of the attractive features of the approach we have given here is that the Schr\"{o}dinger and Pauli Clifford algebras nest in each other showing how naturally the non-relativistic spin model becomes the non-relativistic model without spin simply by going to the sub algebra.  It remains to show that the Dirac particle is described by a larger Clifford algebra which takes into account relativistic effects.  This work will be presented in a later paper \cite{bhbc081}.



\end{document}